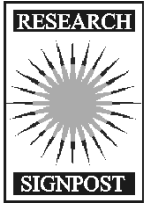

# 2 | Wide-angle x-ray diffraction theory versus classical dynamical theory

**S.G. Podorov[1] and A. Nazarkin[2]**
[1]School of Physics, Monash University, Victoria 3800, Australia
[2]University of Erlangen-Nuremberg, D-91058 Erlangen, Germany

## Abstract

*In this article we provide a comparison of classical dynamical x-ray diffraction theory with the dynamical theory for the wide-angle case. It is shown that it is possible for the true value of the angular variable to be introduced without application of the dispersion theory. Wide-angle x-ray diffraction theory is in an excellent agreement with Zaus correction of the angular parameter.*

Correspondence/Reprint request: Dr. S.G. Podorov, School of Physics, Monash University, Victoria 3800, Australia, University of Erlangen-Nuremberg, D-91058 Erlangen, Germany. E-mail: webmaster@x-ray-soft.de

# 1.Introduction

In this article, we outline a disagreement of the classical dynamical theory of X-ray diffraction with the wide-angle x-ray dynamical diffraction theory. This disagreement follows from different ways to introduce the angular parameters of the X-ray diffraction theory. An alternative way to describe X-ray scattering on deformed crystals is proposed, which does not employ dispersion theory.

The first variants of dynamical X-ray diffraction theory were proposed by Darwin [2] and Ewald [6]. To develop the theory Darwin divided the crystal into thin lamellae, and with the use of suitable recurrence equations he obtained an expression for the reflection coefficient of an ideal crystal. On the basis of the Maxwell theory Ewald (see [6]) developed a system of equations for the amplitudes of both propagated and scattered waves. As the crystal has periodical properties, he expanded the polarizability into a Fourier series, seeking the solution of the scattering problem also in the form of a Fourier series using waves with different wavevectors. Using the periodical properties of both the crystal and the solution, he obtained the system of equations ([1] [Eq. (5.4)]) that is the basis of the classical dynamical theory. The Laue equation [8] in his derivations was taken as dogma valid for all directions of angle of incidence. Following Ewald, the wavevector of the diffracted wave changes its direction and absolute value according to the Laue equation (see [8]). Nevertheless the absolute value of the wavevector of the incident wave remains the same. This point of view on the diffraction theory was not experimentally prooved and is taken as an axiom. Taupin [15] and Takagi [14] extended the theory to the case of crystals with slight deformations. Nevertheless, Taupin and Takagi used a variation of the dispersion theory, i.e they analysed the length of the wavevectors satisfying the Laue equation.

Caticha [4] developed an improved Laue dynamical theory based on a quartic equation for the dispersion surface. This improved theory extends the applicability of Laue theory to the case of the far tails of the Bragg peaks for ideal flat crystals. Further to this work, Caticha [5] extended the dynamical theory of Darwin [2] to include the whole angular range from 0° to 90°. But the correctness of their results was not experimentally prooved. De Caro et al. [3] generalized Laue dynamical theory for strained multilayers. They wrote the theory using a matrix approach that is suitable for computer simulations, but very difficult for mathematical analysis of, for example, solution to the corresponding inverse problem. Even with its later generalizations, the Laue theory is unable to describe reciprocal space mapping for crystals with 3 D deformation or crystals with nano-objects. Grundmann and Krost [7] tested currently-available commercial dynamical theory simulation programs in comparison with atomistic kinematical theory. They demonstrated disagreement of the results from commercial programs with the kinematical theory, for superlattices with large strain.

Zaus [17] showed experimentally that the classical expression for the angular parameter needs correction. Contrary to classical dynamical x-ray diffraction theory, Podorov and Förster [11] developed the dynamical theory without using dispersion theory. Further extension of the theory was made by Podorov et al. [12]. Several ways to introduce the angular variable to diffraction theory are discussed by Podorov and Nazarkin [13].

## 2. Classical x-ray dynamical theory

In the two-beam approximation the equations of classical dynamical theory for incident $E_0$ and reflected $E_g$ waves may be written in following form:

$$2iK_0^z \frac{d}{dz} E_0 + E_0\left[(k^2 - \mathbf{K}_0^2) + k^2\chi_0\right] + k^2 C E_g \chi_{-g} \exp(i\mathbf{g}\mathbf{u}(\mathbf{r})) = 0, \tag{1}$$

$$2iK_g^z \frac{d}{dz} E_g + E_s\left[(k^2 - \mathbf{K}_g^2) + k^2\chi_0\right] + k^2 C E_0 \chi_g \exp(-i\mathbf{g}\mathbf{u}(\mathbf{r})) = 0, \tag{2}$$

in common notation. $\chi_g$ denotes the Fourier components of the crystal susceptibility [9], **g** denotes a diffraction vector, **u(r)** denotes atomic displacement from their ideal position in ideal crystal lattice, and C is a polarization factor.

Eqs. (1-2) are well known Takagi-Taupin type equations (see [14-15]). In the given equations the angular variable is not presented in clear form. To introduce the angular dependence, many authors build the so called dispersion theory.

Let us consider this case in more detail for Bragg symmetrical geometry.

Many authors proposed that the vector $\mathbf{K}_o$ is directed under angle of incidence to the crystal surface and has fixed length $\frac{2\pi}{\lambda}$. The vector $\mathbf{K}_g$ is changing in length and direction following the Laue equation:

$$\mathbf{K}_g = \mathbf{K}_o + \mathbf{g}. \tag{3}$$

Building together with the vectors $K_0$ and **g**. one obtains the "ribbon triangle". Then the angular variable must be introduced in following form:

$$\eta = \frac{\mathbf{K}_g^2 - k^2}{2K_g^z} \tag{4}$$

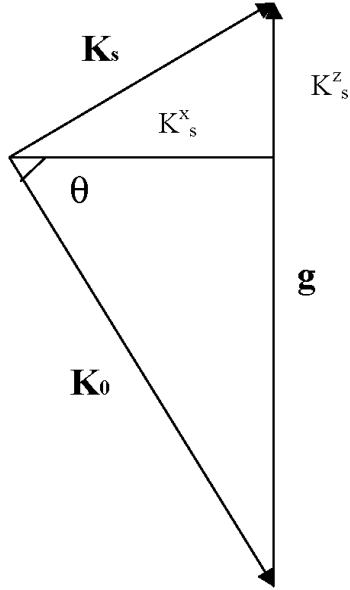

**Figure 1.** Classical treatment of dispersion theory. $\mathbf{K}_0$-wave vector of propagated wave, $\mathbf{K}_g$ is wave vector of reflected wave, $\mathbf{g}$ is diffraction vector and $\theta$ is the angle of incidence. Vectors $\mathbf{K}_0$, $\mathbf{K}_g$ and $\mathbf{g}$ build a triangle to satisfy the Laue equation.

We make exact calculations of the stretching or "dispersion" of vector $\mathbf{K}_g$ (see fig 1):

$$K_g^z = -(g - K_o \sin\theta) = -\frac{2\pi}{\lambda}(2\sin\theta_B - \sin\theta) \tag{5}$$

$$K_g^x = K_0 \cos\theta = \frac{2\pi}{\lambda}\cos\theta \tag{6}$$

$$|\mathbf{K}_g|^2 = |K_0^x|^2 + |K_g^z|^2 = \frac{4\pi^2}{\lambda^2}\cos^2\theta + \left(\frac{4\pi\sin\theta_B}{\lambda} - \frac{2\pi\sin\theta}{\lambda}\right)^2 \tag{7}$$

so by "dispersion theory" the angular variable is exactly

$$\eta = \frac{\mathbf{K}_g^2 - k^2}{2K_g^z} = \frac{\left(\frac{4\pi\sin\theta_B}{\lambda}\right)^2 - 2\left(\frac{4\pi\sin\theta_B}{\lambda}\frac{2\pi\sin\theta}{\lambda}\right)}{-\frac{4\pi}{\lambda}(2\sin\theta_B - \sin\theta)} = \frac{4\pi}{\lambda}\left(\frac{\sin\theta_B \sin\theta}{2\sin\theta_B - \sin\theta} - \sin\theta_B\right) \tag{8}$$

The factor $\dfrac{\sin\theta_B}{2\sin\theta_B - \sin\theta}$ Caticha [4] called an "asymmetry factor"

## 3. Wide-angle dynamical x-ray diffraction theory

The equations of wide-angle dynamical x-ray diffraction theory were introduced by Podorov and Förster [11] and Podorov *et.al.* [12]. The details, of how the angular variable is introduced in the theory, were discussed by Podorov and Nazarkin [13]. The starting equations of the theory are a little different from Eqs. (1-2). We are not using Eq. (3) as "dogma", rather assuming that

$$|\mathbf{K}_g| = |\mathbf{K}_0| = \frac{2\pi}{\lambda} \tag{9}$$

Then:

$$2i(\mathbf{K}_0, \nabla)\mathbf{E}_0 + \mathbf{E}_0 k^2 \chi_0 + k^2 \mathbf{E}_s \chi_{-g} \exp[i(\mathbf{sr} - \mathbf{gr} + \mathbf{gu}(\mathbf{r}))] = 0, \tag{10}$$

$$2i(\mathbf{K}_S, \nabla)\mathbf{E}_S + \mathbf{E}_S k^2 \chi_0 + k^2 \mathbf{E}_0 \chi_g \exp[-i(\mathbf{sr} - \mathbf{gr} + i\mathbf{gu}(\mathbf{r}))] = 0 \tag{11}$$

where s is scattering vector $\mathbf{s} = \mathbf{K}_g - \mathbf{K}_0$. To come to a one-dimensional system of equations we make following substitution:

$$\mathbf{E}_S(\mathbf{r},\mathbf{s}) = \tilde{\mathbf{E}}_S(z,\mathbf{s})\exp(-i(\mathbf{s}-\mathbf{g})\mathbf{r}) \tag{12}$$

$$2iK_0^z \frac{d}{dz}E_0 + E_0 k^2 \chi_0 + k^2 C\tilde{E}_s \chi_{-g} \exp(i\mathbf{gu}(\mathbf{r})) = 0 \tag{13}$$

$$2iK_S^z \frac{d}{dz}\tilde{E}_S + \tilde{E}_s[2\mathbf{K}_S \cdot (\mathbf{s}-\mathbf{g}) + k^2 \chi_0] + k^2 CE_0 \chi_g \exp(-i\mathbf{gu}(\mathbf{r})) = 0 \tag{14}$$

The angular variable then will have following expression

$$\eta = -2\frac{\mathbf{K}_S \cdot (\mathbf{s}-\mathbf{g})}{2K_s^z} = \frac{2\frac{2\pi}{\lambda}\sin\theta \cdot \left(2\frac{2\pi}{\lambda}\sin\theta - \frac{2\pi}{d_0}\right)}{2\frac{2\pi}{\lambda}\sin\theta} = \frac{4\pi}{\lambda}(\sin\theta - \sin\theta_B) \tag{15}$$

Note, that in this case the direction of the reflected wave vector is different from the "dispersion" case.

The expression (15) is in excellent agreement with the value obtained empirically by Zaus [17].

## 4. Numerical comparison of two theories

According Eqs.(1-2) and Eqs. (13-14) we make simulations of the diffraction on the InSb/InGaSb/InSb/InAs SL with the same parameters. We show, that exact calculations following dispersion theory lead to sufficient differences between the two considered theories. As the wide-angle dynamical x-ray diffraction theory is in excellent agreement with Zaus results [17], we may conclude, that the so called "dispersion" theory leads to wrong results. Methodologically the non-dispersion theory is easier to understand and it does not need 100 pages (see Authier [1], or Pinsker [10]) to clarify, how to introduce angular dependence in the diffraction theory.

## 5. Conclusion

As we showed, there are no reasons to assume that the wave vectors and diffraction vector build a "ribbon triangle". The dispersion theory is built on the dogma of the Laue equation. Note that the Laue equation is obtained for ideal semi-infinite crystals and is not valid for any cases. Conversely, the non-dispersion theory gives better agreement with experimental data.

In conclusion, we showed disagreement of the classical dynamical x-ray diffraction theory with results obtained by Zaus [17] and the wide-angle x-ray diffraction theory.

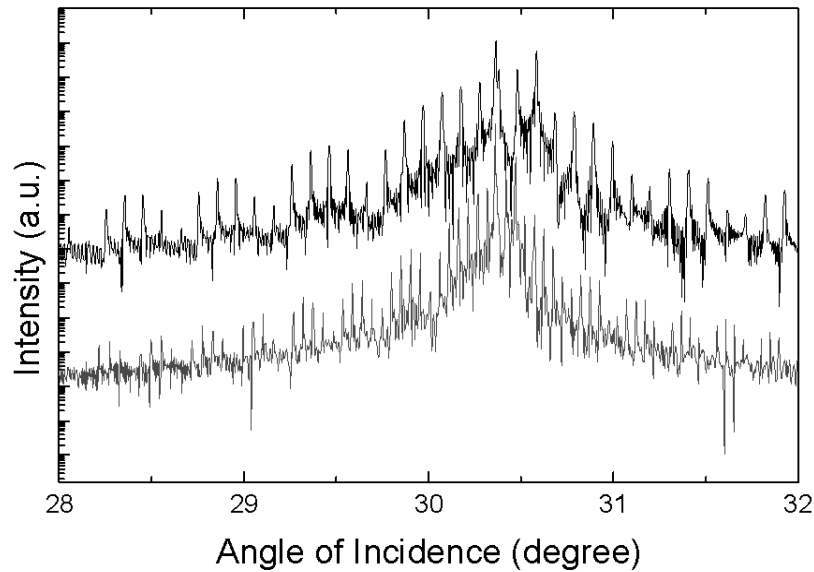

**Figure 2.** Simulated diffracted intensity in logarithmic scale from a InSb/ InGaSb/ InSb/ InAs SL a)- by wide-angle x-ray diffraction theory (above, black) b) - by classical x-ray diffraction theory with dispersion of diffracted wavevector (below, magenta, shifted for clarity).